\documentstyle[floats,multicol,aps,epsf,prb]{revtex}
\newcommand{\sgn}{\mathop{\rm sgn}\nolimits}
\begin{document} 
\draft
\twocolumn[\hsize\textwidth\columnwidth\hsize\csname @twocolumnfalse\endcsname

\title{ Large$-N$ treatment of the Abrikosov transition at low temperatures.}
\author{A. Lopatin, G. Kotliar}
\address{Department of Physics, Rutgers University, Piscataway, 
New Jersey  08855}
\date{\today}
\maketitle

\begin{abstract}

We investigate the influence of order parameter fluctuations
on the transition between  normal and  mixed superconducting states
at low temperatures. We show that in case of clean  quasi-two-dimensional 
superconductors the transition  can be described by
the functional of the Ginzburg-Landau type. We consider the 
large$-N$ generalization of this functional
and  using the lowest Landau level
approximation we get the large$-N$ equations which describe the phase
transition. In case of physical dimensionality
we  found that  the transition is of the first order. The fluctuations 
significantly
affect  the temperature dependence of the upper critical field.  
  
\end{abstract}

\vskip2pc]

\section{ Introduction .}

It is well known that  the magnetic field   penetrates  type$-II$
superconductors  through  flux lines which form the Abrikosov
lattice \cite{1}.
 The theory of this mixed state was first developed by Abrikosov 
for temperatures close to $T_c$ \cite{1},  and then it was 
extended to all temperatures\cite{hw}.  These  theories ignore the 
fluctuations of the order parameter. It is a very good approximation
for the conventional superconductors because in this case 
the order parameter
fluctuations are important only in a very small region 
near the phase transition line. This  happens because  the Ginsburg
numbers of the conventional superconductors are very small.
 
However for   high$-T_c$ superconductors there are experiments which
cannot be  explained by the usual mean field theory.
The upper critical field $H_{c2}$ at low temperatures significantly increases
as temperature decreases  instead of being approximately
constant as follows from the mean field theory\cite{2}.
 The Ginzburg numbers of high$-T_c$ superconductors
are not very small, therefore one can suggest that this
unusual behavior is due to the order parameter
fluctuations.  Also, it is important
to understand the type of the phase transition (first order  or
second order), because in  mean field approximation this
transition is always of the second order, but the fluctuations
can induce the first order phase transition. This
happens, for example, in the model  described in Ref.\cite{braz}
and it was  suggested to  happen in the Abrikosov
transition \cite{brezin}. 

  In the following we will consider the influence of the order parameter
fluctuations on the transition between normal and mixed superconducting 
states  in clean  superconductors.
We argue that in case of pure quasi-two-dimensional superconductors,
when the magnetic field is applied along the low-conducting
direction,
even at low temperatures, this transition can be investigated by
the effective functional of the Ginzburg-Landau (GL) type which contains
imaginary  time, because quantum fluctuations become  important at low
temperatures. 
Unfortunately,  even having the  effective functional
one can hardly calculate the free energy exactly, which   is typical
for the critical phenomena problems. 
Therefore we modify the functional
introducing $n-$ index to the fluctuating field and considering the large$-N$
limit.
 One should be careful in introducing of $n-$ index into
 this functional, because doing that in not a proper way
one can get a  model which does not have 
a solution in the form of the Abrikosov lattice\cite{moore}.
 Under the approximation when the mean field solution and fluctuations
belong to the lowest Landau level (which is  valid
near the phase transition) we get the  equations 
which describe the phase transition.

We show that in high enough
dimensions (either $ T=0,\, d>4$ or $T\not=0,\, d>6$) if the coupling constant
is not too large  the transition 
is of the second order.
In this  case the corrections from fluctuations
do not  modify mean field results essentially.
 In the physical case (d=3) the phase transition is always of the first
order and the fluctuations significantly affect the phase transition line:
 the upper critical field increases as
temperature decreases, and the curvature of
this dependence is negative when the temperature is not too low
(see later, Figs.4,5).

 The paper is organized as follows: In Sec. \ref{2} we describe the model.
In Sec. \ref{3} we derive the large-$N$ equations describing
the phase transition. In Sec. \ref{4}
we analyze the phase transition from the side of the normal state
region. In Sec. \ref{5}
we simplify the large-$N$ equations making the lowest Landau level
approximation. In Sec. \ref{6} we solve the large$-N$  equations
in case of high dimensions. In Sec. \ref{7}  we solve
the large-$N$ equations in the physical case. In Sec. \ref{8} 
we discuss the spectrum of the order parameter fluctuations.
 And we summarize our results in Sec. \ref{9}.

\section{ Description of the model.} \label{2}

 To analyze the transition to  the superconducting state one can
usually use GL  functional.  This functional
is based on the  expansion in the order parameter, so it is valid  when the 
ratio of the order parameter and temperature ${\Delta/T}$ is small.
In case when an external magnetic field  close to  $H_{c2}$ is applied,
one can still  use  GL functional if ${\Delta/T}\ll 1$, but at
low temperatures the expansion in this parameter becomes not possible.
From the other hand  the  magnetic field produces a
depairing effect, as temperature does, so one can try to
expand in $\Delta$ because it is small compared with the magnetic field taken
in the proper units. This  was done in Ref.\cite{maki} and it was shown that
the coefficient in ${|\phi|}^4$ term  has a logarithmic singularity
$\ln{1\over T},$ which means that the expansion in $\Delta$
is not possible at zero temperature. The appearance of this
singularity  can be easily  understood physically:
Indeed, if we consider a pair
of  electrons which move  exactly parallel to the magnetic field, then the 
magnetic field does not affect them  in the quasiclassical approximation.
 Therefore this singularity
comes from the electrons moving parallel to  the magnetic field or close
to this direction, because 
the magnetic field does not produce a  depairing effect
 for these electrons.
The direct calculation of the coefficient of  ${|\phi|}^4$ term
shows, that indeed, the singularity comes from the momenta
parallel to the magnetic field.

 There is a way to avoid this problem if one considers a
quasi-two-dimensional superconductor. Indeed, if the magnetic
field is applied along the low-conducting direction, then due to the 
quasi-two-dimensional band structure there are no momenta parallel
to the magnetic field. In this case it is possible to expand 
in $\Delta$ and get a functional of GL type. Unfortunately,
the functional which arises does not match exactly
the form of the usual GL functional. The difference is that
the ${|\phi|}^4$ term becomes a  nonlocal functional of $\phi$
fields with a range of ``interaction'' of the order of the 
magnetic length\cite{maki}.
We hope that this difference is not crucial, and therefore
we consider the model with a local ${|\phi|}^4$ term.
The Lagrangian density of this model is
\begin{eqnarray}
{\cal L}=\phi^*\left(-{{D^2}\over{2m}}+\epsilon(\hat p_\perp)
+|\partial_\tau|+a\right)\phi
+u{|\phi|}^4   \nonumber \\
+{{B^2}\over{8\pi}}
-{{BH}\over{4\pi}} \label{lagden},
\end{eqnarray}
where $D_\mu=\nabla_\mu-{{ie^*}\over c}A_\mu$ acts only in $x-y$ plane with
the Abrikosov lattice, and  $\epsilon_\perp={{p_\perp^2}\over{2m_\perp}}$ 
is the
kinetic energy corresponding to the motion in the directions perpendicular 
to $x-y$ plane.
 In this Lagrangian $\phi$ is a function of coordinates and 
imaginary time because we want to consider the low-temperature
quantum regime also.
The operator $|\partial_\tau|$ means  that in the Matsubara space this
operator becomes $|\omega|.$
For the BCS model the coefficients in the Lagrangian (\ref{lagden}) are
\begin{equation}
u\sim{1\over{p_F m_e}},\,\,
{1\over{m_\perp}}\sim{{v_F^2}\over{\epsilon_0 k_a^2}}
,\,\,\, a+\epsilon_0\sim\epsilon_0 {{H-H_{c2}^{(0)}}\over{H_{c2}^{(0)}}}
\label{coeff},
\end{equation}
where $\epsilon_0$ is the energy of the lowest Landau level, i.e. the
lowest eigenvalue of ${{D^2}\over{2m}},$  $m_e$ is the electron mass,
 $H_{c2}^{(0)}$ is the critical field in the mean field approximation,
and $k_a=t_{\|}/t_{\perp}$ is the anisotropy coefficient, which  is
the ratio of the in-plane electron hopping and the hopping in 
the direction perpendicular 
to the $x-y$ planes. 

To make our approach systematic we will modify this model
introducing $n-$index and considering  the large$-N$ limit.
The usual way to introduce $n-$index to ${|\phi|}^4$ term  is
\begin{equation}
\phi^*\phi\phi^*\phi\to\phi^*_n\phi_n\phi^*_m\phi_m. \label{largen}
\end{equation}
But it happens that
 the spectrum of fluctuations
around the Abrikosov lattice is not positive definite in this case, i.e.
the Abrikosov lattice is  unstable\cite{moore}.
(We will not consider the possibility of condensation into different
n-components as  in Ref.\cite{moore}.)  To understand the reason of
this instability
let us substitute $\phi=\phi_0+\phi_1,$  where $\phi_0$
corresponds to the Abrikosov lattice and $\phi_1$ represents
fluctuations around it, into  ${|\phi|}^4$ term. 
Considering the simplest case when fluctuations are small, for 
 ${|\phi|}^4$ term we have
\begin{equation}
{|\phi|}^4=4\phi_0^*\phi_0\phi_1^*\phi_1+\phi_0^*\phi_0^*\phi_1\phi_1+
\phi_0\phi_0\phi_1^*\phi_1^*+... \label{eilen},
\end{equation}
where we wrote only the most interesting, quadratic in $\phi_1$ part.
The spectrum of fluctuations corresponding to (\ref{eilen})
was considered  by Eilenberger  \cite{eilen}, and it was found that it is 
 positive  definite, i.e. the Abrikosov lattice is stable.
It is important that following to the Eilenberger approach
one can see that  these are the off-diagonal terms in (\ref{eilen})
which make the lattice stable. But for 
${|\phi|}^4$ term defined by (\ref{largen})
in the large$-N$ limit we have
$$\phi^*_n\phi_n\phi^*_m\phi_m=...+2\phi^*_0\phi_0\phi_m^*\phi_m +...,$$
so in this case we effectively drop out the  off-diagonal terms and it leads
 to the unstable spectrum of fluctuations.
 Therefore we suggest the following modification of the model:
\begin{equation}
\phi^*\phi\phi^*\phi \to 2\phi^*_n\phi_n\phi^*_m\phi_m
-\phi^*_n\phi_n^*\phi_m\phi_m. \label{largen1}
\end{equation}
In the large$-N$ limit for ${|\phi|}^4$ term now we have
$$
4\phi_0^*\phi_0\phi_n^*\phi_n-\phi^*_0\phi^*_0\phi_m\phi_m-
\phi_0\phi_0\phi^*_m\phi^*_m,
$$
which matches the  form (\ref{eilen}) after the redefinition 
$\phi_m\to i\phi_m,\phi_m^*\to -i\phi_m^*.$ It is important that the 
second term in
(\ref{largen1}) should have the negative sign  
because otherwise the 
fluctuations around the Abrikosov lattice are unstable
(from  the above simple
argument one cannot see that it should be negative).
 So the Lagrangian density of the model which we will consider is:
\begin{eqnarray}
{\cal L}=\phi^*_n\left(-{{D^2}\over{2m}}+\epsilon(p_\perp)
+|\partial_\tau|+a\right)\phi_n
+{{B^2}\over{8\pi}}-{{BH}\over{4\pi}} \nonumber  \\
+{u\over N}(2\phi^*_n\phi_n\phi^*_m\phi_m 
-\phi^*_n\phi_n^*\phi_m\phi_m) \label{lag},
\end{eqnarray}

and  the action and the partition function are
\begin{eqnarray}
S=\int d\tau d^d x {\cal L},\\
Z=\int D\phi D\phi^* e^{-S}.
\end{eqnarray}
 For simplicity, in the following   we will neglect the fluctuations
of the magnetic field, because in the lowest Landau level approximation
it gives just a renormalization of $u$ term, \cite{eilen,affleck}
moreover the high-$T_c$ superconductors are extremely type-II 
superconductors therefore even this renormalization is not essential.

\section{ Large$-N$ equations.} \label{3}

The interaction term in the Lagrangian (\ref{lag}) can be decoupled 
with the  help of a real field  $\rho$ and a complex one  $\Delta$
\begin{eqnarray}
 {\cal L}=\phi^*_n\left({\cal E}
+|\partial_\tau|+a+2\rho i\right)\phi_n
+\Delta^*\phi_n\phi_n+\Delta\phi^*_n\phi^*_n  \nonumber  \\  
+{{B^2}\over{8\pi}}-{{BH}\over{4\pi}} 
+{N\over{2u}}\rho^2+{N\over u}\Delta^*\Delta,
\end{eqnarray}
where ${\cal E}=-{{D^2}\over{2m}}+\epsilon(p_\perp)$ is
the kinetic energy operator.
 From the form of this Lagrangian it is evident that the mean field value 
of $\rho$ should be imaginary, which  is not a problem because one can always
shift the  contour of integration over $\rho$ in the complex plane.
Therefore we simply redefine  $\rho\to -i\rho$, because we will not consider
the fluctuations of $\Delta$ and $\rho$ fields.
Integrating  over $N-1$ $\phi$ fields we get an effective Lagrangian
\begin{eqnarray}
{\cal L}&=\phi^*_0\left({\cal E}
+|\partial_\tau|+a+2\rho \right)\phi_0 
+\Delta^*\phi_0\phi_0+\Delta\phi_0^*\phi_0^*& \nonumber \\
&-{N\over{2u}}\rho^2+{N\over u}\Delta^*\Delta 
+{{N-1}\over 2}Tr \ln G^{-1}
 +{{B^2}\over{8\pi}}-{{BH}\over{4\pi}}& \nonumber,
\end{eqnarray}
\vspace{0.5cm}
where the Green function $G$
satisfies the equation
$$
\left[\begin{array}{cc}
{\cal E}+|\omega|+a+2\rho& 2\Delta  \\
2\Delta^*&{\cal E}^*+|\omega|+a+2\rho
\end{array}\right]G({\bf r}_1,{\bf r}_2,\omega,p_\perp) 
$$
\begin{equation}
\hspace{5cm}=\delta({\bf r}_1-{\bf r}_2) \label{g}
\end{equation}

 Taking the   variation with respect to $\Delta,\rho,\phi_0$ and neglecting
fluctuations of the magnetic field we get the large$-N$ equations
$$
\left[ \begin{array}{cc}
{N\over 2u}\rho({\bf r})-\phi_0^*({\bf r})\phi_0({\bf r}) 
& -{N\over u}\Delta({\bf r})-\phi_0({\bf r})\phi_0({\bf r}) \\
-{N\over u}\Delta^*({\bf r})-\phi_0^*({\bf r})\phi_0^*({\bf r})
&{N\over 2u}\rho({\bf r})-\phi_0^*({\bf r})\phi_0({\bf r}) \end{array} \right]
$$
\begin{equation}
\hspace{3cm}={{(N-1)T}\over V_\perp}
\sum_{\omega,p_\perp}G({\bf r},{\bf r},\omega,p_\perp),
\label{syst1}
\end{equation}
\begin{equation}
\left(-{{D^2}\over{2m}}+a+2\rho({\bf r})\right)\phi_0({\bf r})
+2\Delta({\bf r})\phi_0^*({\bf r})=0,
\label{syst2}
\end{equation}
where $V_\perp$ denotes the volume perpendicular to the planes with the 
Abrikosov lattice.

   \section{ Normal state.} \label{4}

 At temperatures higher than the temperature of the Abrikosov transition
we have $\phi_0=0,\Delta=0$, and the equations (\ref{syst1},\ref{syst2}) 
are reduced to
\begin{equation}
{1\over{2u}}\rho= {T\over{V_\perp}}\sum_{\omega,p_\perp}G(0,0,\omega,p_\perp),
\end{equation}
where the limit $N\to\infty$ was taken.
The Green function in this case is determined by 
\begin{equation}
\left({\cal E}+a+2\rho+|\omega|\right)
G({\bf r},0,\omega,p_\perp)=\delta^2({\bf r}).
\end{equation}
 Under the approximation when $\phi$ field belongs to the lowest Landau level
we can substitute $-{1\over{2m}}D^2=\epsilon_0 $, where $\epsilon_0$ is the
energy of the lowest Landau level and get
\begin{equation}
{\rho\over{2u}}={{TH}\over{V_\perp\Phi_0}}\sum_{\omega,p_\perp}
{1\over{\delta+|\omega|+\epsilon(p_\perp)}}e^{-{1\over{\tilde\epsilon}}
(\delta+\epsilon(p_\perp)+|\omega|)} \label{tc},
\end{equation}
where  $\epsilon_0+2\rho+a=\delta$ and $\Phi_0={{2\pi c}\over{e^*}}.$
Also we  introduced the ultra-violate cutoff $\tilde\epsilon.$ 
The advantage of this cutoff procedure is that the Green function with the
cutoff can be written as
\begin{eqnarray}
{ 1 \over{\delta+|\omega|+\epsilon(p_\perp)}}
&e^{-{1\over{\tilde\epsilon}} (\delta+|\omega|+\epsilon(p_\perp))} \nonumber \\
=\int_{1\over{\tilde\epsilon}}^{\infty}&{{d\lambda}\over\lambda}
e^{-\lambda(\delta+\epsilon(p_\perp)+|\omega|)}.
\end{eqnarray}

The phase transition happens when the correlation length diverges, i.e. when 
$\delta\to 0$,
therefore to find the phase transition line
we should calculate  the integrals in (\ref{tc}) for the 
case $T\gg\delta$ in which we
have
$$
\hspace{-3cm}\delta-a^*-\epsilon_0
={{4uH}\over{\Phi_0}}\left({{m_\perp}
\over{2\pi}}
\right)^{{d_{\perp}}\over 2} 
$$
\begin{equation}
\hspace{1cm} \times  \left\{
\begin{array}{ccc}
{\pi\over{3(d_\perp /2-2)}}{\tilde\epsilon}^{d_\perp/2-2}\,\,T^2
,\,\,\, d_\perp >4 \vspace{0.4cm} \\
\pi^{d_{\perp}/2-1}\,\xi\,\, T^{d_\perp/2},  \,\,\,\,2<d_{\perp}<4
\vspace{0.4cm} \\
 \Gamma(1-d_{\perp}/2)\, \delta^{d_{\perp}/2-1}\,\,T  ,\,\,\,\,d_{\perp}<2,
\end{array}\right. \label{trline}
\end{equation}
where $a^*$  is the ``mass''  term $a$ renormalized by the quantum 
fluctuations 
\begin{equation}
a^*=a+{{8uH}\over{\Phi_0\pi d_\perp}} \left({{m_\perp {\tilde
\epsilon}}\over{2\pi}}\right)^{d_\perp/2},  \label{quantren}
\end{equation}
 $\xi$ is a constant
$$\xi=\int_0^\infty{{d\lambda}\over{\lambda^{d_\perp/2}}}
(\coth\lambda-{1\over\lambda}),$$
and we introduced a notation for the dimensions perpendicular to $x-y$ planes
$d_\perp=d-2.$
One can see that in high dimensions  $d_\perp>4$ the correction
to the phase transition line is analytical in $T,$
therefore it does not change the mean field
result essentially. When $2<d_\perp<4$ 
the correction is non-analytical.
Substituting $a^*+\epsilon_0\sim\epsilon_0(H-H_{c2}^{(0)*})/H_{c2}^{(0)},$ 
for the last case we have 
$$ H_{c2}-H_{c2}^{(0)*}\sim - T^{d_\perp/2},\,\,\,\, 2<d_{\perp}<4, $$
where $H_{c2}^{(0)*}$ is the mean field upper critical field 
renormalized by the quantum fluctuations (see (\ref{quantren})).
In the physical case $d_{\perp}=1$ one can see that the r.h.s of
(\ref{trline}) diverges as ${1\over{\sqrt{\delta}}}$ when $\delta\to 0.$
 In this case  (\ref{trline}) can be written in the form
\begin{equation}
{{\delta-a^*-\epsilon_0}\over{\epsilon_0}}=2\kappa_G {T\over{\epsilon_0}}
\sqrt{{{\epsilon_0}\over\delta}},\label{kappa}
\end{equation}
where $\kappa_G$ is the Ginzburg number
\begin{equation}
\kappa_{G}={{2uH}\over{\sqrt{\epsilon_0}\Phi_0}}
\left({{m_\perp}\over 2}\right)^{1\over 2} \label{ginzburg}.
\end{equation}
Using (\ref{coeff}) we can estimate 
$\kappa_G\sim{{k_a}\over{p_F^2S}}$, where $S$ is the
area corresponding to  the unit flux $\Phi_0.$ By the order of magnitude
$S\sim\xi^2$, where $\xi$ is the coherence  length.
In case when $\kappa_G$ is small one can find the qualitative crossover
line between the Gaussian region (where the u-term is not important)
and the non-Gaussian region  (when it becomes important).
In the Gaussian regime one can neglect the r.h.s. in (\ref{kappa})
getting $\delta\approx a^*+\epsilon_0.$ The crossover line
corresponds to the situation  when the r.h.s of (\ref{kappa})
becomes of the same order with the l.h.s., so that  we have
$\delta\sim (\sqrt{\epsilon_0}\kappa_G T)^{2/3}$ or
\begin{equation}
{{H_{cr}-H_{c2}^{(0)*}}\over{H_{c2}^{(0)}}}\sim -\left({{\kappa_G T}\over
{\epsilon_0}}\right)^{2/3}, \label{astim}
\end{equation}
where $H_{cr}$ is the magnetic field corresponding to the crossover
from the Gaussian to the non-Gaussian  regime.
The fact that one cannot reach the ordered state going from the normal state
gives us  a hint that the transition to the ordered state
 is not continues.

\section{ Ordered state.} \label{5}

 We will solve Eqs.(\ref{g},\ref{syst1},\ref{syst2}) under 
the approximation when the 
mean field $\phi_0$ and the fluctuations around it belong to the lowest
Landau level. Also, as usually, we will consider a lattice with the triangular
symmetry. Therefore, following to the Eilenberger notations\cite{eilen}, we 
take $\phi_0$ to be proportional to
\begin{equation}
\phi({\bf r}|0)=(2\eta)^{1\over 4}
\sum_{p}e^{{{2\pi}\over\eta}\left(-{1\over 2}(y+p\eta)^2
+ip\eta(x+{1\over 2}p\xi)\right)} \label{abr},
\end{equation}  
where $\eta,\xi$ are components of  the vectors which determine the unit cell
\begin{equation}
{\bf r}_I=(1,0)
,\,\,\,{\bf r}_{II}=(\xi,\eta)=({1\over 2},{{\sqrt 3}\over 2}).
\end{equation}
 In the above formulas we measure all distances in the distances 
between the vertices
$l$, which is related  to the magnetic field by 
\begin{equation}
\Phi_0=H{{\sqrt 3}\over 2} l^2.
\end{equation} 
Applying the magnetic translation operator to (\ref{abr}) we can get a 
complete basis
of   functions belonging   to the lowest Landau level
\begin{equation}
\phi({\bf r}|{\bf r}_0)=e^{2\pi i{{y_0 }\over \eta} x} 
\phi({\bf r}+{\bf r}_0|0), \label{tr}
\end{equation}
which obey the  orthogonality relation
\begin{equation}
\int d^2 r\,\phi^*({\bf r}|{\bf r_1})\phi({\bf r}|{\bf r}_2)=
\eta^2\delta^2({\bf r}_1-{\bf r}_2). 
\end{equation}
 
 Restricting our solution to the lowest Landau level we artificially
decrease the space of solutions of Eqs.(\ref{g},\ref{syst1},\ref{syst2}),
 therefore to have a solution
we should decrease the space of equations projecting them on  the
lowest Landau level. The way to do that becomes evident if we 
consider the Schroedinger equation which corresponds to  the Green 
function equation (\ref{g})
\begin{equation}
\left(-{{D^2}\over{2m}}+\epsilon(p_\perp)+|\omega|+a+2\rho\right)\phi
+2\Delta\phi^*=E\phi. \label{seq}
\end{equation}
Indeed, to project this equation on the lowest Landau level
one should multiply it by $\phi^*({\bf r}|{\bf r}_0)$ and 
integrate over ${\bf r}.$
Introducing the following notations for the matrix elements 
of $\rho$ and $\Delta$
\begin{eqnarray}
\int d^2 r \, \phi^*({\bf r}|{\bf r}_1)\rho({\bf r})\phi({\bf r}|{\bf r}_2)
=\eta^2\rho_{{\bf r}1,{\bf r}2}, \\
\int d^2 r \, \phi^*({\bf r}|{\bf r}_1)\Delta({\bf r})\phi^*({\bf r}|{\bf r}_2)
=\eta^2\Delta_{{\bf r}1,{\bf r}2},
\end{eqnarray}
and presenting $\phi$ as
\begin{equation}
\phi({\bf r})=\int d^2 r_0 a({\bf r}_0)\phi({\bf r}|{\bf r}_0),
\end{equation}
we get
\begin{eqnarray}
\left(\epsilon_0+\epsilon(p_\perp)+|\omega|+a\right) a({\bf r}_1)& \nonumber \\
+2\int d^2 r_2\, a({\bf r}_2)&(\rho_{{\bf r}1,{\bf r}2}
+\Delta_{{\bf r}1,{\bf r}2})=E \,a({\bf r}_1).\nonumber
\end{eqnarray}
 From the symmetry of this problem we expect that $\rho$ is 
a periodic function, and
$\Delta$ is a quasiperiodic one (that means  periodic up to the phase), 
therefore the only matrix elements which are allowed
by the translational symmetry are
\begin{equation}
\rho_{{\bf r}1,{\bf r}2}=\rho_{{\bf r}1}\delta^2({\bf r}_1-{\bf r}_2), 
\end{equation}
\begin{equation}
\Delta_{{\bf r}1,{\bf r}2}=\Delta_{{\bf r}1}\delta^2({\bf r}_1+{\bf r}_2).
\end{equation}
\newpage
 Now  we can find the eigenfunctions and the energy levels of (\ref{seq}):
\begin{eqnarray}
u_{\pm}({\bf r})={1\over{\sqrt 2}}
e^{i{{\theta(r)}\over 2}}\left(e^{ikz}\phi({\bf r}|{\bf r}_0)
\pm e^{-ikz}\phi({\bf r}|-{\bf r}_0)\right),  \label{egu} \\
v_{\pm}({\bf r})={1\over{\sqrt 2}}
e^{i{{\theta(r)}\over 2}}\left(ie^{ikz}\phi( {\bf r}|{\bf r}_0)
\mp ie^{-ikz}\phi({\bf r} |-{\bf r}_0)\right), \label{egv}
\end{eqnarray}
\begin{equation}
E_\pm({\bf r},p_\perp,\omega)
=\epsilon_0+\epsilon_\perp(p_\perp)+a+2(\rho_{\bf r}\pm |\Delta_{\bf r}|),
\end{equation}
where
$$
e^{i\theta({\bf r})}={{\Delta_{\bf r}}\over{|\Delta_{\bf r}|}}.
$$
 One can see that there are two energy branches which are  denoted by
the subscript $\pm$, and there are two eigenfunctions $u$ and $v$
for each energy level.

The green function (\ref{g}) can be expressed through the eigenfunctions
(\ref{egu},\ref{egv})

\begin{eqnarray}
&G({\bf r}_1,{\bf r}_2,p_\perp,\omega)& \nonumber \\
&={1\over{2\eta^2 l^2}}
\sum_{s,j}&\int^\prime d^2 r_0{{\Psi_{j,s}({\bf r}_1|{\bf r}_0)
\Psi_{j,s}^\dagger
({\bf r}_2|{\bf r}_0)}\over{E_j({\bf r}_0,p_\perp,\omega)}},\label{guv}
\end{eqnarray}
where the index $j=+,-$ denotes the spectrum branch and $s=u,v$ 
denotes the type of function:
\begin{eqnarray}
\Psi_{\pm,u}=\left(\begin{array}{cc} u_\pm \\ u^*_\pm \end{array}\right),
\Psi_{\pm,u}^\dagger=\left(u_\pm^*\,\, u_\pm\right), \\
\Psi_{\pm,v}=\left(\begin{array}{cc} v_\pm \\ v^*_\pm \end{array}\right),
\Psi_{\pm,v}^\dagger=\left(v_\pm^*\,\, v_\pm\right).
\end{eqnarray}
 The prime under the integral in (\ref{guv})
denotes the integration over the half of the unit cell.
 From the Goldstone theorem we expect  a singularity in (\ref{guv})
at the minimal energy, i.e. $E_-(0,0,0)=0.$ This condition is  consistent with 
the equation (\ref{syst2}) if  we  take 
$$\phi_0=i \sqrt{bN}\phi({\bf r}|0),$$
where $b$
is a real positive number. Indeed, projecting (\ref{syst2}) on the 
lowest Landau level
we get 
\begin{equation}
\epsilon_0+a+2(\rho_0-\Delta_0)=0,
\end{equation}
which is the same with $E_-(0,0,0)=0.$

Now, to get a closed system of equations, we should take the  matrix
elements of the equation (\ref{syst1}). Using the addition theorem of the 
Eilenberger 
functions \cite{eilen}.

$$
\phi({\bf r}|{\bf r}_1)\phi({\bf r}|{\bf r}_2)={1\over{\sqrt{2}}}\biggl[
\tilde\phi({\bf r}|({\bf r}_1+{\bf r}_2)/2)
\tilde\phi(({\bf r}_1-{\bf r}_2)/2|0)\biggr.
$$
$$
+\tilde\phi({\bf r}|({\bf r}_1+{\bf r}_2+{\bf r}_I)/2)
\tilde\phi(({\bf r}_1-{\bf r}_2)/2|{\bf r}_I/2)\biggr] ,
$$
where $\tilde\phi$ is defined by (\ref{abr},\ref{tr})
 with the difference $\tilde {\bf r}_{II}
={1\over 2}{\bf r}_{II}$, we get
$$
-{T\over{2\eta^2 V_\perp l^2}}\sum_{p_\perp,\omega,j}
\int^\prime d^2 r_0 {{\sgn j \,\, e^{i\theta(r_0)}}
\over{E_{j}({\bf r}_0, p_\perp,\omega)}}
\biggl[\tilde\phi^*({\bf r}|0)\tilde\phi({\bf r}_0|0)
$$
\begin{equation}
+ \tilde\phi^*({\bf r}|{\bf r}_I/2)
\tilde\phi({\bf r}_0|{\bf r}_I/2)\biggr]
={{\Delta_{\bf r}}\over u}-b Q({\bf r})        \label{mdel}
\end{equation}
$$
-{T\over{2\eta^2 V_\perp l^2}}\sum_{p_\perp,\omega,j}\int^\prime d^2 r_0
{{K({\bf r}-{\bf r}_0)+K({\bf r}+{\bf r}_0)}
\over{E_{j}({\bf r}_0, p_\perp,\omega)}}
$$
\begin{equation}
\hspace{4cm}=-{{\rho_{\bf r}}\over u}+b K({\bf r}), \label{mrho}
\end{equation}
where 
\begin{equation}
\sgn j=\left\{\begin{array}{c} 1 \,\,\,\, {\rm  for \, (+)\, branch } \\
 -1 \,\,\,\,  {\rm for \, (-)\,  branch } \end{array}\right. ,
\end{equation}
and  we introduced the following functions
\begin{eqnarray}
K({\bf r})=\left|\tilde\phi({\bf r}/2|0)\right|^2+
\left|\tilde\phi({\bf r}/2| {\bf r}_I/2)\right|^2,  \\
Q({\bf r})={1\over 2}\left(
\tilde\phi^*({\bf r}|0)\tilde\phi(0|0)+
\tilde\phi^*({\bf r}|{\bf r}_I/2)\tilde\phi(0|{\bf r}_I/2)
\right).
\end{eqnarray}

 From Eq.(\ref{mdel}) one can see that $\Delta_{\bf r}$ is a linear combination
of the functions $\tilde\phi^*({\bf r}|0)$ and  $\tilde\phi^*
({\bf r}|{\bf r}_I/2).$ Therefore, presenting $\Delta_{\bf r}$ as
\begin{equation}
\Delta_{\bf r}={1\over 2}\left(\Delta_1
\tilde\phi^*({\bf r}|0)\tilde\phi(0|0)+\Delta_2
\tilde\phi^*({\bf r}|{\bf r}_I/2)
\tilde\phi(0|{\bf r}_I/2) \label{prdel}
\right),
\end{equation}
from (\ref{mdel}) we get two equations on $\Delta_1,\Delta_2$
\begin{eqnarray}
{T\over{\eta^2 V_\perp l^2}} \sum_{p_\perp,\omega,j}\int d^2 r_0 {{\sgn j\, 
e^{i\theta(r_0)}}
\over{E_{j}({\bf r}_0, p_\perp,\omega)}}
\tilde\phi({\bf r}_0|0) \nonumber \\
=\tilde\phi(0|0)\left(b-{{\Delta_1}\over u}\right) \label{ld1} 
\end{eqnarray}
\begin{eqnarray}
{T\over{\eta^2 V_\perp l^2}} \sum_{p_\perp,\omega,j}\int d^2 r_0 {{\sgn j\, 
e^{i\theta(r_0)}}
\over{E_{j}({\bf r}_0, p_\perp,\omega)}}
\tilde
\phi({\bf r}_0|{\bf r}_I/2) \nonumber \\
=\tilde\phi(0|{\bf r}_I/2)
\left(b-{{\Delta_2}\over u}\right) \label{ld2}
\end{eqnarray}
 From the symmetry of the problem we expect that $\Delta_{\bf r}$ is not only
quasiperiodic (which is evident from (\ref{prdel})), but it is also
quasisymmetric under rotations on $\pi/3.$ The later property 
is satisfied only 
when $\Delta_1=\Delta_2.$ But taking  $\Delta_1=\Delta_2$ we 
should prove that Eqs.(\ref{ld1},\ref{ld2}) become  linear dependent.
One can check that it happens indeed, multiplying (\ref{ld1}) by
$\tilde\phi(0|{1\over 2}{\bf r}_I)$ and (\ref{ld2}) by 
$-\tilde\phi(0|0)$, adding them, and using that the function
\begin{eqnarray}
D\equiv\left(\tilde\phi^*({\bf r}|0)\tilde\phi(0|0)+
\tilde\phi^*({\bf r}|{1\over 2}{\bf r}_I)\tilde\phi(0|{1\over 2}{\bf r}_I)
\right) \nonumber \\
\times \left(\tilde\phi({\bf r}|0)\tilde\phi(0|{1\over 2}{\bf r}_I)-
\tilde\phi({\bf r}|{1\over 2}{\bf r}_I)\tilde\phi(0|0)
\right)
\end{eqnarray} 
transforms under rotation on $\pi/3$  as
\begin{equation}
D\to D\,\, e^{-i{{2\pi}\over 3}} \,\,\,\, {\rm when}\,\,
x+iy\to (x+iy) e^{i\pi/3}.
\end{equation}
 
Finally we get the following system of equations:
\begin{eqnarray}
{T\over{2\eta^2 V_\perp l^2}} \sum_{p_\perp,\omega,j}\int d^2 r{{\sgn j}
\over{E_{j}({\bf r}, p_\perp,\omega)}}|Q({\bf r})| \nonumber \\
=\left(b-{\Delta\over u}\right)Q(0),   \label{eq1} 
\end{eqnarray}
\begin{eqnarray}
{T\over{2\eta^2 V_\perp l^2}} \sum_{p_\perp,\omega,j}\int d^2 r_0{ 1
\over{E_{j}({\bf r}_0, p_\perp,\omega)}}K({\bf r}_0-{\bf r}) \nonumber \\
={{\rho_{\bf r}}\over u}-bK({\bf r}),         \label{eq2}
\end{eqnarray}
\begin{equation}
E_\pm({\bf r},p_\perp,\omega)=\epsilon_0+a+\epsilon(p_\perp)+|\omega|
+2\left(\rho_{\bf r}\pm\Delta|Q({\bf r})|\right),       \label{eq3}
\end{equation}
\begin{equation}
E_-(0,0,0)=0.        \label{eq4}
\end{equation}

\section{ Solution of the large$-N$  equations in high dimensions.} \label{6}

 In high enough dimensions we can expand  Eqs.(\ref{eq1},\ref{eq2}) 
 in $\rho,\Delta.$ This expansion is  possible when either 
$d_\perp>4$, $T\not= 0$ or
$d_\perp>2$, $T=0.$ In these cases one can write
\begin{equation}
{{Tu}\over{2\eta^2 V_\perp l^2}}\sum_{p_\perp,\omega}
{1\over{\epsilon(p_\perp)+|\omega|+x}}=\alpha-\beta x.\label{expansion}
\end{equation}
 From  Eqs.(\ref{eq1},\ref{eq2}) we get
\begin{equation}
-4\beta\Delta\int d^2r\,\,Q^*({\bf r})Q({\bf r})=(bu-\Delta)Q(0) \label{hd1},
\end{equation}
\begin{eqnarray}
2\int d^2 r_0\,\,\left(\alpha-\beta(\epsilon_0+a+2\rho({\bf r}_0))\right)
K({\bf r}_0-{\bf r}) \nonumber \\
=\rho_{\bf r}-bu K({\bf r}). \label{hd2}
\end{eqnarray} 
 The first equation (\ref{hd1}) can be solved immediately giving
\begin{equation}
\Delta={{bu}\over{1-2\beta\eta}} \label{hd3}.
\end{equation} 
 The second equation (\ref{hd2}) can be solved by the Fourier 
transformation which we
define as 
\begin{equation}
K({\bf r})=\sum_k e^{-i{\bf k}{\bf r}}K({\bf k}),
\end{equation}
where due to the periodicity of $K({\bf r})$ the vector ${\bf k}$ should 
have the 
following discrete values
\begin{eqnarray}
{\bf k}={\bf k}_1 n_1+{\bf k}_2 n_2, \\
{\bf k}_1=2\pi(0,-{2\over{\sqrt{3}}}) \,\,\,\, ,
{\bf k}_2=2\pi(1,-{1\over{\sqrt{3}}}) \label{k1k2},
\end{eqnarray}
where $n_1,n_2$ are integers. The Fourier transformation of $K({\bf r})$
can be done  analytically
\begin{equation}
K({\bf k})=e^{-{{2\pi}\over{\sqrt{3}}}(n_1^2+n_1n_2+n_2^2)}.
\end{equation}
Defining the Fourier transformation for $\rho_{\bf r}$ by the same rule,
from Eq.(\ref{hd2}) we get
\begin{equation}
\rho({\bf k})={{4(\alpha-\beta(\epsilon_0+a))
\eta\,\,\delta_{{\bf k},0}+bu K({\bf k})}
\over{1+4\beta\eta K({\bf k})}} \label{hdr}.
\end{equation}
 Substitution of  (\ref{hd3},\ref{hdr}) to (\ref{eq4}) gives the 
equation for the condensate density $b$
\begin{equation}
2bu\left(
{{Q(0)}\over{1-2\beta\eta}}-\sum_k{{K({\bf k})}\over{1+4\eta\beta K({\bf k})}}
\right)=
{{\epsilon_0+a+8\eta\alpha}\over{1+8\eta\beta}}.\label{cond}
\end{equation}
Using this equation we can present the spectrum as
\begin{equation}
E_\pm({\bf r})=bu\left(\sum_k { {K({\bf k}) {\,\,e^{i {\bf k}{\bf r}}} } 
\over{1+4\beta\eta K({\bf k})}} 
\pm{{|Q({\bf r})|}\over{1-2\beta\eta}}\right)-({\bf r}=0). 
\label{hdspectr}
\end{equation}
 One can check that this spectrum is
positive definite which means that our solution is stable.
 The transition point can be found  from (\ref{cond}) taking $b=0$
\begin{equation}
\epsilon_0+a_c=-8\eta\alpha,
\end{equation}
where $a_c$ means the critical value for $a.$
 The same result can be obtained from (\ref{tc}). We expect to
have a nonzero $b$ in the ordered state when $a<a_c.$
But  Eq.(\ref{cond}) has positive solutions  for $b$ 
(and $b$ was chosen to be positive)
only when the expression in the parenthesis on
the l.h.s. of (\ref{cond}) is negative.
Numerical calculation of this expression gives
that it is negative only when
\begin{equation}
\beta<\beta_c\label{con},
\end{equation}
where $\beta_c=0.112.$ When the condition (\ref{con}) is
satisfied the condensate density  $b$ can have any small values 
as  a function of $a$.
Therefore in that case the phase transition is of the second order.
Note that the expansion (\ref{expansion}) is valid only
when $\Delta,\rho,b\ll\epsilon_0.$ When the phase transition is
of the second order this condition can be always satisfied
if we close enough to the phase transition line. Therefore the
absence of the solution when $\beta>\beta_c$ in fact means that the
transition is not continues.

Let us now summarize the results of this section where we were interested in 
the cases
 $d>4, T\not=0$ and $d>2, T=0$ : When the interaction constant is not
too strong (condition (\ref{con})) 
the phase transition is of
the second order. If the condition (\ref{con}) is not satisfied
then the transition between the normal to the ordered states  cannot be
continues.

\section{ Physical case } \label{7}

Let us consider the physical case $d=3$ ($d_\perp=1$).
Taking the integral over $ p_\perp$ and summing  over 
$\omega$ in (\ref{eq1},\ref{eq2})
we get
\begin{eqnarray}
t\int d^2 r\,\, \biggl[\sqrt{T} \Bigl(f(\epsilon_+({\bf r})/T) 
-f(\epsilon_-({\bf r})/T)\Bigr) \nonumber \\
-{2\over{\sqrt{\pi}}}\left(\sqrt{\epsilon_+({\bf r})}
-\sqrt{\epsilon_-({\bf r})}\right)
\biggr] |Q({\bf r})|   \nonumber  \\
=(bu-\Delta)Q(0)  
\end{eqnarray}
\begin{eqnarray}
t\int d^2 r_0 \,\,\biggl[\sqrt{T} \Bigl(f(\epsilon_+({\bf r}_0)/T) 
-f(\epsilon_-({\bf r}_0)/T)\Bigr) \nonumber \\
-{2\over{\sqrt{\pi}}}\left(\sqrt{\epsilon_+({\bf r}_0)}
-\sqrt{\epsilon_-({\bf r}_0)}\right)
\biggr] K({\bf r}_0-{\bf r}) \nonumber \\
=\rho_{\bf r}-bu K({\bf r})-{{4\sqrt{3\tilde\epsilon}}\over\pi}t,
\end{eqnarray}
where 
\begin{equation}
\epsilon_{\pm}({\bf r})=\epsilon_0+a+2(\rho_r\pm\Delta|Q({\bf r})|), 
\end{equation}
\begin{equation}
f(x)=\int_0^\infty {{d\lambda}\over{\sqrt{\lambda}}}
e^{-\lambda x}\left(\coth\pi\lambda-{1\over{\pi\lambda}}\right),
\end{equation}
and
\begin{equation}
t={{u\sqrt{m_\perp}}\over{2 \eta^2 l^2\sqrt{2\pi}}}.
\end{equation}
 It is convenient to introduce the following dimensionless variables
\begin{equation}
T=\epsilon_0\kappa_G^2 T^\prime ,\,\,\,
\Delta=\epsilon_0\kappa_G^2 \Delta^\prime,   
\end{equation}
\begin{equation}
\rho=\epsilon_0 \kappa_G^2 \rho^\prime , \,\,\,
bu=\epsilon_0\kappa_G^2 b^\prime, 
\end{equation}
\begin{equation}
\epsilon_0+a=\epsilon_0 \kappa_G^2 a^\prime,
\end{equation}
where $\kappa_G$ is the Ginzburg number (\ref{ginzburg}). 
In these new variables
the equations do not contain $\kappa_G$ explicitly:
\begin{eqnarray}
k\int d^2 r\,\, \biggl[\sqrt{T^\prime} 
\Bigl(f(\epsilon_+^\prime({\bf r})/T^\prime) 
-f(\epsilon_-^\prime({\bf r})/T^\prime)\Bigr) \nonumber \\
-{2\over{\sqrt{\pi}}}\left(\sqrt{\epsilon_+^\prime({\bf r})}
-\sqrt{\epsilon_-^\prime({\bf r})}\right)
\biggr] |Q({\bf r})|  \nonumber \\
=(b^\prime-\Delta^\prime)Q(0),  
\end{eqnarray}
\begin{eqnarray}
k\int d^2 r \,\,\biggl[\sqrt{T^\prime} 
\Bigl(f(\epsilon_+^\prime({\bf r}_0)/T^\prime) 
-f(\epsilon_-^\prime({\bf r}_0)/T^\prime)\Bigr) \nonumber \\
-{2\over{\sqrt{\pi}}}\left(\sqrt{\epsilon_+^\prime({\bf r}_0)}
-\sqrt{\epsilon_-^\prime
({\bf r}_0)}\right)
\biggr] K({\bf r}_0-{\bf r})  \nonumber \\
=\rho_{\bf r}^\prime-b^\prime K({\bf r})
-{{4\sqrt{3\tilde\epsilon}}\over{\pi\epsilon_0}}t, \label{meq2}
\end{eqnarray}
where $k=1/(2\sqrt{3\pi})$ is a numerical constant.
Note that the third term on the r.h.s. of Eq.(\ref{meq2}) can be absorbed into
renormalization of $\rho^\prime$ and $a^\prime$
$$\rho^\prime\to\rho^\prime+ {{4\sqrt{3\tilde\epsilon}}\over{\pi\epsilon_0}}t$$
$$a^\prime\to a^\prime-{{8\sqrt{3\tilde\epsilon}}\over{\pi\epsilon_0}}t. $$
This  corresponds to the renormalization of the external magnetic field
by the quantum  fluctuations as in (\ref{quantren}).
 Eq.(\ref{meq2}) is an integral equation with the kernel
\begin{equation}
K({\bf r})=\sum_{n_1,n_2} e^{-{{2\pi}\over{\sqrt{3}}}(n_1^2+n_1 n_2+n_2^2)}
e^{-2\pi i({\bf k}_1 n_1+{\bf k}_2 n_2){\bf r}} \label{kernel},
\end{equation}
where ${\bf k}_1,{\bf k}_2$ were defined in (\ref{k1k2}).
To simplify the problem we will take an approximate expression
for this kernel leaving only zeroth and  first harmonics:
\begin{eqnarray}
K({\bf r})=2+\gamma\biggl[\cos 2\pi{{2y}\over{\sqrt{3}}}+
\cos 2\pi\left(x+{y\over{\sqrt{3}}}\right) \nonumber \\
+\cos 2\pi\left(x-{y\over{\sqrt{3}}}\right)\biggr],
\end{eqnarray}
where $\gamma=4 e^{-2\pi/\sqrt{3}}\approx 0.1063.$ It seems to
be a good approximation because the contribution from  higher
harmonics to (\ref{kernel}) decreases 
exponentially with the harmonic order. For example the contribution from
the second harmonic is less then  $ 0.1 \% $ of the first one. Presenting
$\rho_{\bf r}^\prime$ in the form
\begin{eqnarray}
\rho_{\bf r}^\prime=\rho_0+\gamma\rho_1\biggl[\cos 2\pi{{2y}\over{\sqrt{3}}}+
\cos 2\pi\left(x+{y\over{\sqrt{3}}}\right) \nonumber \\
+\cos 2\pi\left(x-{y\over{\sqrt{3}}}\right)\biggr],
\end{eqnarray}
we get the following system of equations

\newpage
\begin{eqnarray}
2k\int d^2 r\,\, \biggl[\sqrt{T^\prime} 
\Bigl(f(\epsilon_+^\prime({\bf r})/T^\prime) 
+f(\epsilon_-^\prime({\bf r})/T^\prime)\Bigr) \nonumber \\
-{2\over{\sqrt{\pi}}}\left(\sqrt{\epsilon_+^\prime({\bf r})}
+\sqrt{\epsilon_-^\prime({\bf r})}\right)\biggr] 
=\rho_0-2b^\prime,    \label{fsq1}    
\end{eqnarray}
\begin{eqnarray}
k\int d^2 r \,\,\biggl[\sqrt{T^\prime} 
\Bigl(f(\epsilon_+^\prime({\bf r})/T^\prime) 
+f(\epsilon_-^\prime({\bf r})/T^\prime)\Bigr) \nonumber \\
-{2\over{\sqrt{\pi}}}\left(\sqrt{\epsilon_+^\prime({\bf r})}
+\sqrt{\epsilon_-^\prime
({\bf r})}\right)
\biggr] \cos 2\pi{{2y}\over{\sqrt{3}}} 
=\rho_1-b^\prime,  \label{xib} 
\end{eqnarray}
$$
k\int d^2 r\,\, \biggl[\sqrt{T^\prime} 
\Bigl(f(\epsilon_+^\prime({\bf r})/T^\prime) 
-f(\epsilon_-^\prime({\bf r})/T^\prime)\Bigr) 
$$
\begin{equation}
-{2\over{\sqrt{\pi}}}\left(\sqrt{\epsilon_+^\prime({\bf r})}
-\sqrt{\epsilon_-^\prime({\bf r})}\right)
\biggr]|Q({\bf }r)|  
=(b^\prime-\Delta^\prime)Q(0), \label{bdel} 
\end{equation}
\begin{equation}
\epsilon_{\pm}^\prime({\bf r})=2\Bigl(\rho_1\gamma\,\, s(x,y)
+\Delta^\prime (Q(0)\pm|Q({\bf r})|)\Bigr) \label{spect}
\end{equation}
\begin{equation}
\rho_0=\Delta^\prime  Q(0)-3\rho_1 \gamma-a^\prime /2 \label{rhozero}
\end{equation}
where 
\begin{eqnarray}
s(x,y)=\cos 2\pi{{2y}\over{\sqrt{3}}}+
\cos 2\pi\left(x+{y\over{\sqrt{3}}}\right) \nonumber \\
+\cos 2\pi\left(x-{y\over{\sqrt{3}}}\right)-3.
\end{eqnarray}
 Eq.(\ref{rhozero}) follows from Eq.(\ref{eq4}). Note that the spectrum 
(\ref{spect}) contains only
$\Delta^\prime$ and $\rho_1$, therefore combining  
Eqs.(\ref{xib},\ref{bdel})  to 
get rid of $b^\prime$ on the right hand sides  we  get a closed equation on 
$\Delta^\prime$ and  $\rho_1.$
Therefore taking a given $\Delta^\prime$ we can solve this equation for  
$\rho_1.$
Then, knowing $\Delta^\prime$ and $\rho_1$ we can find all the other
parameters $b^\prime,a^\prime,\rho_0.$
 In Figs. 1-3  we present the graphs of $\rho_1,b^\prime,a^\prime$
as  functions of $\Delta^\prime$ for different temperatures.
Note that for large $\Delta^\prime$ the fluctuation contribution
is small and one should have the mean field results:
\begin{equation}
b^\prime=\rho_1=\Delta^\prime,\,\,\,a^\prime=-2Q(0)b^\prime,\,\,\,\, 
\Delta^\prime\to\infty.
\end{equation}
One can see that the graphs on Figs.1-3 begin to approach this asymptotics.

Let us start the analysis of these graphs from the low-temperature case
(see Fig. 1):
One can see that $a^\prime$ first increases as  $\Delta^\prime$ decreases  
(as one should expect from the mean field theory), but then it decreases. 
The extremum point corresponds to the first order phase transition.

\begin{figure}[here]
\begin{center}
\epsfxsize=9.0 cm
\epsfysize=8.0 cm
\epsfbox{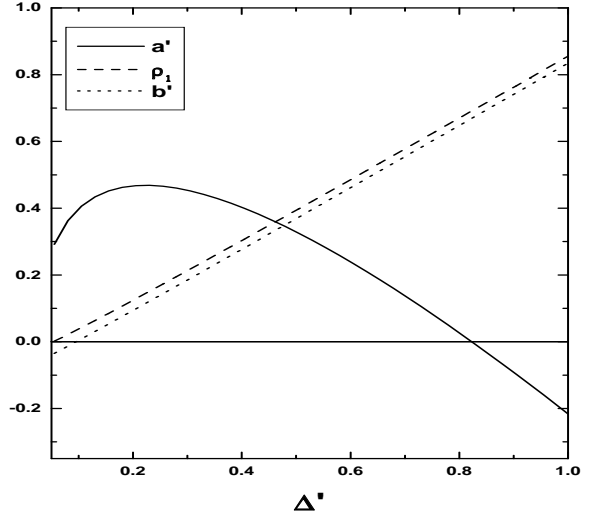}
\end{center}
\caption{Solution of Eqs.(\ref{fsq1}-\ref{rhozero})
at temperature $T^\prime=0.1.$ The extremum in $a^\prime$
corresponds to the first order phase transition. Note that 
the condensate density $b^\prime$ is finite at this point.}
\end{figure}

At high temperatures the behavior is different (see Fig. 3): 
Going from large  $\Delta^\prime$
we see that in this case first the condensate density $b^\prime$ becomes zero,
and then there is an extremum in $a^\prime.$ In our approach it was chosen
that $b^\prime>0,$ therefore the point $b^\prime=0$ corresponds to the 
first order phase transition. We call this transition also first order one 
 because it is still discontinues in $\rho$ and $\Delta.$
So, the transition is of the first order for  all temperatures, but at 
low temperatures
it corresponds to the extremum of $a^\prime$, while at high
temperatures  it corresponds to depleting of the condensate 
density to zero. This change in the kind of the first order phase transition
 takes place at $T^\prime=T^*\approx2.6.$

\begin{figure}[here]
\begin{center}
\epsfxsize=9.0 cm
\epsfysize=8.0 cm
\epsfbox{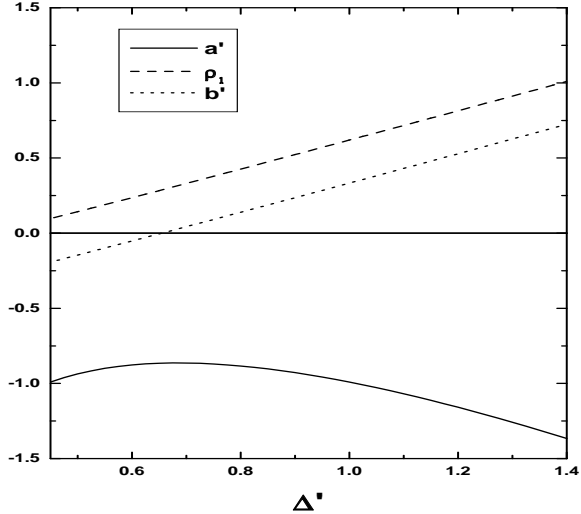}
\end{center}
\caption{Solution of Eqs.(\ref{fsq1}-\ref{rhozero})
at temperature $T^\prime=2.0$ The condensate density $b^\prime$
is close to zero at the point of extremum of $a^\prime.$}
\end{figure}

\begin{figure}[here]
\begin{center}
\epsfxsize=9.0 cm
\epsfysize=8.0 cm
\epsfbox{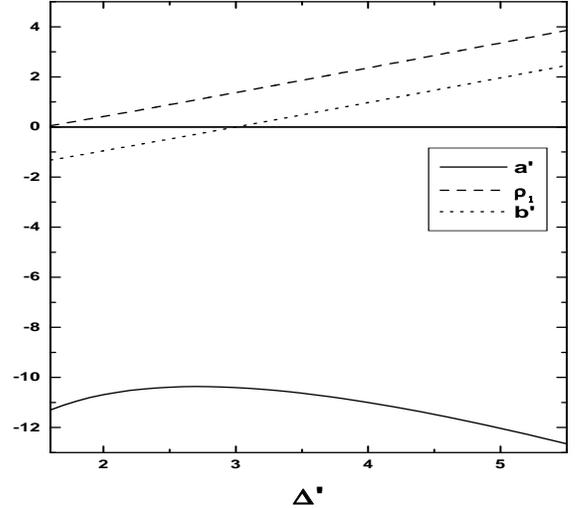}
\end{center}
\caption{Solution of Eqs.(\ref{fsq1}-\ref{rhozero})
at temperature $T^\prime=20.$ The first order phase transition
corresponds to  the point where  $b^\prime=0.$}
\end{figure}

\begin{figure}[here]
\begin{center}
\epsfxsize=9.0 cm
\epsfysize=8.0 cm
\epsfbox{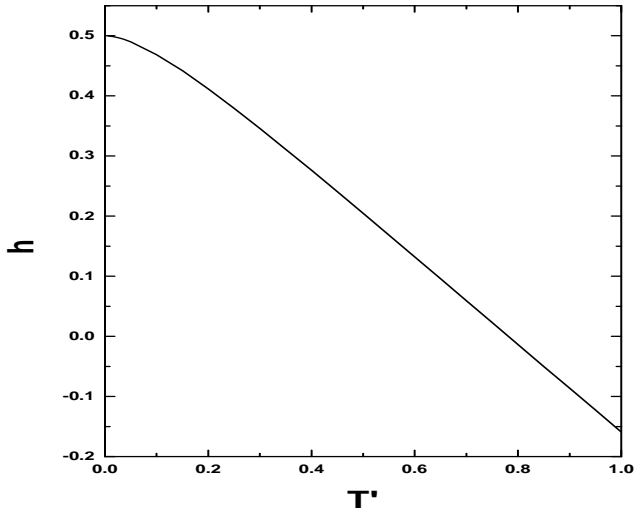}
\end{center}
\caption{First order phase transition line at low temperatures.}
\end{figure}

\begin{figure}[here]
\begin{center}
\epsfxsize=9.0 cm
\epsfysize=8.0 cm
\epsfbox{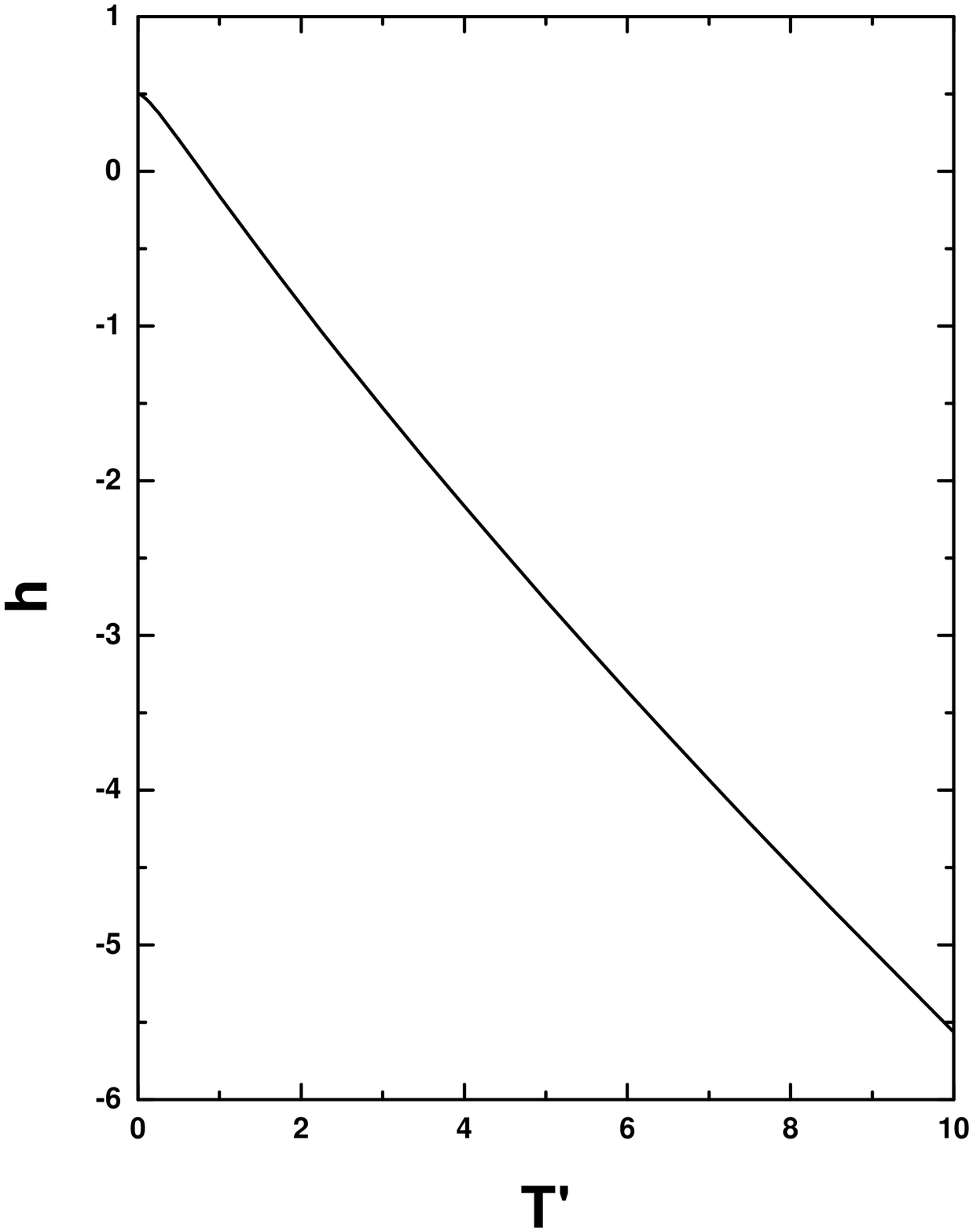}
\end{center}
\caption{First order phase transition line.}
\end{figure}

The phase transition curve which was obtained from
the above criteria is 
\begin{equation}
a_c^\prime(T^\prime)=h(T^\prime),
\end{equation}
where $h$ is plotted on Figs. 4,5.
The asymptotic behavior of the function $h$ at large arguments is
\begin{equation}
h(x)\approx -2.92\,\, x^{2/3}+2.31\,\, x^{1/2},\,\, x\gg 1,
\end{equation}
with an accuracy of leading and sub-leading terms.
At low temperatures the function $h(T^\prime)$ is analytical in $T^\prime.$
In spite of the fact that there is a change in the kind of the transition 
at $T^*,$
this curve does not have a break at this point.
 The phase transition line in the original notations is
\begin{equation}
a_c^*+\epsilon_0=\epsilon_0\kappa_G^2\,\,
h\left({T\over{\epsilon_0\kappa_G^2}}\right) \label{avst},
\end{equation}
where $a^*$ is  $a$ renormalized by quantum fluctuations (\ref{quantren}).
 One can see that this curve scales with the Ginzburg
parameter $\kappa_G$. 
\newpage
According to the form of (\ref{avst}) one can define
too regimes: the low-temperature, quantum regime
\begin{equation}
T\ll\epsilon_0\kappa_G^2,
\end{equation}
 and high-temperature, classical one
\begin{equation}
T\gg\epsilon_0\kappa_G^2.
\end{equation}
 Note that the quantum region extends as $\kappa_G$ increases.
 Asymptotically in the classical regime from (\ref{avst})
one gets 
\begin{equation}
 a_c^*+\epsilon_0=-2.92 \,\epsilon_0
\left({{\kappa_G T}\over{\epsilon_0}}\right)^{2/3},
\end{equation}
which  agrees with the  estimation (\ref{astim}).

\section{ Low-energy spectrum.} \label{8}

 In this section we will show that the low-energy 
spectrum of  fluctuations in our model
is  different from one which  follows from  the 
Eilenberger theory. In our
model  the Eilenberger result can be obtained
if  one neglects the terms on the l.h.s. of
 Eqs.(\ref{eq1},\ref{eq2}). These  terms  are small if we are
not too close to the phase transition line, nevertheless they 
are always not equal to zero, and as we will show they
change the low-energy spectrum considerably.
 If we  neglect  the mentioned terms in  Eqs.(\ref{eq1},\ref{eq2}) then  
we have
\begin{eqnarray}
\Delta=bu,   \\
\rho_{\bf r}=buK({\bf r}), 
\end{eqnarray}
and for the lower energy branch we get
\begin{equation}
E_-({\bf r})=2bu(K({\bf r})-|Q({\bf r})|)-2bu(K(0)-|Q(0)|),
\end{equation}
 where we used that $E_-(0)=0.$
 It happens that if  we expand this expression
in powers  of ${\bf r},$  then the terms quadratic in ${\bf r}$
cancel each other and the expansion starts from 
$r^4$ term
\begin{equation}
E_-({\bf r})\sim r^4.
\end{equation}
 This result leads to  divergences, for example the fluctuation
contribution to the density in case $d=3$ ,$T\neq 0$ is
logarithmically divergent
\begin{equation}
\int {{d^2 r \,dp}\over{p^2+E_-({\bf r})}}\sim\int {{d^2 r}\over{r^2}}.
\end{equation}
Also the fluctuation  correction to the conductivity
has a logarithmic singularity \cite{maki1}.
The fact that there are infra-red divergences means that one needs
a more careful analysis of the infra-red behavior of the model.
The similar situation happens in the two-dimensional Bose gas
at non zero temperatures.  The Bogolubov approximation
leads to the divergent fluctuation contribution to the density,
but the careful analysis of the infrared behavior based
on the effective low-energy functional leads to the 
theory without  divergences. To the best of our knowledge
the asymptotic behavior of the model under consideration was not found yet.

  In our large-$N$  model this problem does not arise because 
due to the fluctuation contribution
the fine tuning
between  the particle-hole and particle-particle parts 
of the spectrum ($\rho_{\bf r}$ and $\Delta| Q({\bf r}|$)
does not happen, and one
expects  that the leading term in the low-energy 
spectrum is $r^2$
\begin{equation}
E_-({\bf r})\sim r^2.
\end{equation}
 In case of high dimensions one can see this explicitly
from  Eq.(\ref{hdspectr}).
 And in general case, one can see that
the particle-particle and particle-hole parts of the
spectrum are affected by the fluctuation
terms in a different way and therefore we do not expect any 
fine tuning between these  terms.

 We think that the result $E_-({\bf r})\sim r^2$ is specific for the 
large$-N$ model, and the situation in  the real model
is much more complicated. Nevertheless, we think that
the large$-N$ model is a reasonable model for
the description of the  phase transition, because the 
infra-red properties  seem to be irrelevant for the  phase transition. 
Indeed, usually the infra-red divergences are absent 
in the perturbation expansion
for the physical quantities like  free energy, density, etc. And it is 
enough to know 
the free energy to determine  the kind of the phase transition and
to  find the phase transition line.

\section{ Discussion and Conclusions.} \label{9}

We considered the effect of order parameter fluctuations on the
transition between  normal and mixed superconducting states
in pure superconductors.
Our starting point was an effective functional of  GL
type. We showed that the coefficients in this  functional
are finite (i.e. this functional exists) in the quasi-two-dimensional
situation, when the applied magnetic field is parallel to the low-conducting
direction.
This case is interesting from the point of view of
high$-T_c$ superconductors, because they have a quasi-two-dimensional
band structure.
 We considered  this functional in the large$-N$ limit. One should be careful
  introducing the n-index into the Lagrangian, because the symmetry 
between the particle-hole and particle-particle channels
 is very important for this problem. Indeed, introducing the n-index
in the  usual way
\begin{equation}
\phi^*\phi\phi^*\phi\to\phi^*_n\phi_n\phi^*_m\phi_m,
\end{equation}
 in the  large$-N$ limit one effectively drops out the  particle-particle 
channel,
and it leads to the model with an unstable spectrum of fluctuations.
Therefore we introduce the n-index in  the following way:
\begin{equation}
\phi^*\phi\phi^*\phi \to 2\phi^*_n\phi_n\phi^*_m\phi_m-
\phi^*_n\phi_n^*\phi_m\phi_m. \label{ph}
\end{equation}   
  The coefficients in the above formula can be found  from the following
consideration: The effect of fluctuations can be formally suppressed
reducing the Ginzburg number.  And in this limiting case the Eilenberger
theory becomes exact. We want our large$-N$ model to be
as close to the real model as possible, and therefore in the 
limiting case $\kappa_G\to 0$ we should have  the Eilenberger answer for the
spectrum. This requirement uniquely defines the coefficients in (\ref{ph}).

 To simplify the large$-N$ equations we used the lowest Landau level 
approximation
which is valid when the order parameter is much smaller than 
$\epsilon_0,$ i.e. when we are not too far from the phase transition line.
These large$-N$ equations can be easily solved in case of high dimensions:
either  $d_\perp>4,\, T\not=0$ or  $d_\perp>2,\, T=0.$ In these cases the 
transition
was found to be  of the second order if the interaction  constant is
not too large. It is interesting
to draw a parallel between our solution of the large$-N$ model
and the renormalization group  approach to the quantum critical
phenomena problems\cite{millis,hertz}. In our case the dynamical 
exponent $z=2.$
Note that the magnetic field ``eats'' two dimensions,
therefore the straightforward application of the 
results \cite{millis,hertz} gives that 
the upper critical dimension at zero temperature is $d_\perp= 4-z=2,$
which agrees with our approach. 
 The phase transition line in the large$-N$ limit was found to be
\begin{equation}
 H_{c2}-H_{c2}^{(0)*}\sim - T^{d_\perp/2} \,\,,\,\,\,2<d_{\perp}<4, 
\label{chd} 
\end{equation}
where $H_{c2}^{(0)*}$ is the mean field upper critical field
renormalized by the quantum fluctuations.
In fact this  result is more general than that of the   large$-N$ limit.
  Indeed, the difference between 
our model and the standard GL  model 
(which was considered in Ref.\cite{millis})  arises
only when one considers the renormalization of $u$ term \cite{brezin}.
 But in the case under consideration  the $u$ term is irrelevant 
and therefore one should get 
the usual answer. Therefore the answer (\ref{chd}) should hold in case
$N=1$ too. 

 In case of physical dimensionality ($d_\perp=1$)  the model gives 
the first order phase transition. The fact that at finite temperatures 
the phase transition is of the first order looks natural
because the fluctuation contribution 
 diverges as one reaches the
phase transition line from the normal state. 
  (For example the first order correction
to the ``mass''  term diverges as $1/\sqrt{\delta}$, see (\ref{trline}).)
This situation is similar  to one which  happens in the model studied by
Brazovskiy in Ref.\cite{braz}, where the fluctuations drive the  
phase transition 
to the first order one. Therefore we think that in the real model
at finite temperatures the transition is  of the first order too.
 At zero temperature the large$-N$ model also gives the first order
phase transition, but it is not clear whether  in the real model the 
transition should be necessarily of the first order at zero
temperature.

  The phase transition line which follows from our model is
\begin{equation}
a_c^*+\epsilon_0=\epsilon_0\kappa_G^2\,\,
h\left({T\over{\epsilon_0\kappa_G^2}}\right),
\end{equation}  
where $h$ is plotted on Figs. 4,5, and 
\begin{equation}
{{a_c^*+\epsilon_0}\over{\epsilon_0}}
\sim {{H_{c2}-H_{c2}^{(0)*}}\over{H_{c2}^{(0)}}} \label{ans}.
\end{equation}
Note that we considered only the low-temperature part of the phase
diagram $T\ll\epsilon_0,$ so that the result (\ref{ans}) may  be applied  only
in this case.
 According to the form of  (\ref{ans}) one can define to regimes:
$T\ll\epsilon_0 \kappa_G^2$ corresponding  to the quantum fluctuations,
and $T\gg\epsilon_0 \kappa_G^2$ corresponding  to the classical ones.
Note that increase of the Ginzburg number makes the problem more quantum.
For example if $\kappa_G\sim 1$, then one cannot reach the classical 
region because
our theory works when  $T\ll\epsilon_0.$
Asymptotically in the classical region the phase transition line is
\begin{equation}
a_c^*+\epsilon_0=-2.92 \,\epsilon_0
\left({{\kappa_G T}\over{\epsilon_0}}\right)^{2/3}.
\end{equation}
 Qualitatively the phase transition line looks similar to the 
experimental data \cite{2} on overdoped high$-T_c$ materials: The upper 
critical
field significantly increases as temperature decreases showing a 
nonanalytical dependence.
 In our theory, the curvature of the phase transition line is negative in the 
classical regime, but at low temperatures it becomes positive (see Fig.4). 
 Note that the Ginzburg
number in  this problem is proportional to the anisotropy 
$\kappa_G\sim{{k_a}\over{p_F^2S}}$, that enhances the fluctuation
contribution. Indeed the resistive phase transition is broad in this
materials, that  supports that the fluctuation contribution is large.

Finally, to avoid confusion, we note that in the high temperature
superconductors, a line in the H-T plane   referred to    
as the irreversibility line,    is usually  interpreted in terms of the
melting   of the vortex lattice. To address these experiments
our  work should be generalized to incorporate the effects of disorder
 which are beyond the scope of our paper. 
However, low enough disorder should not affect 
the melting line. Therefore in that case the transition
line which was found in the paper can be considered as
the  irreversibility line.

\end{document}